\documentclass[12pt,letterpaper]{article}
\usepackage[utf8]{inputenc}
\usepackage[version=3]{mhchem} 
\usepackage{siunitx} 
\usepackage{graphicx} 
\usepackage{natbib} 
\usepackage{amsmath} 
\usepackage{graphicx}
\usepackage[outdir=./]{epstopdf}
\usepackage{natbib}
\usepackage{multirow}
\usepackage{xcolor}
\usepackage{newtxtext,newtxmath}
\usepackage{graphicx,subfigure}
\usepackage{multirow}
\usepackage{tabulary,booktabs}
\usepackage{lineno}
\usepackage{amsmath}
\usepackage{hyperref}
\usepackage{xspace}
\usepackage{array}

\title{Zernike moments description of solar and astronomical features: Python code} 
\author{Hossein \textsc{Safari}$^{1,2}$\thanks{E-mail: safari@znu.ac.ir}, Nasibe \textsc{Alipour}$^{3}$\thanks{E-mail: nasibealipour@gmail.com}, Hamed \textsc{Ghaderi}$^{1}$ and  Pardis \textsc{Garavand}$^{1}$ }

\date{$^{1}$Department of  Physics, Faculty of Science, University of Zanjan, University Blvd., 45371-38791, Zanjan, Iran\\%
$^{2}$Observatory, Faculty of Science, University of Zanjan, University Blvd., 45371-38791, Zanjan, Iran\\%
$^{3}$ Department of Physics, University of Guilan, Rasht, 41335-1914, Iran; \\
} 

\begin{document}

\maketitle
\section{Abstract} 
Due to the massive increase in astronomical images (such as James Web and Solar Dynamic Observatory), automatic image description is essential for solar and astronomical. Zernike moments (ZMs) are unique due to the orthogonality and completeness of Zernike polynomials (ZPs); hence valuable to convert a two-dimensional image to one-dimensional series of complex numbers. The magnitude of ZMs is rotation invariant, and by applying image normalization, scale and translation invariants can be made, which are helpful properties for describing solar and astronomical images. In this package, we describe the characteristics of ZMs via several examples of solar (large and small scale)  features and astronomical images. ZMs can describe the structure and morphology of objects in an image to apply machine learning to identify and track the features in several disciplines.

\section{Introduction}
Objects recognition in images has been developed in several disciplines \citep[e.g.,][]{ref:goshtasby1985, ref:heywood1995, ref:aschwanden2010image,ref:zheng2015improved,MORADKHANI2015123,Noori2019}. Recently, feature extraction for machine learning of object finding and tracking based on image moments was investigated \citep{honarbakhsh}. Moments are a class of image description  \citep[e.g.,][]{ref:hu1962visual, ref:teh1988image}. 

Since the image data of various fields such as biology, medicine, optics, astronomy, and solar physics are vastly recorded, these images' descriptions are out of manual analysis. 
The image moments are quantities that describe an image's shape, objects, and structure. 

Zernike moments (ZMs) map an image to a complex number series \citep{book,1057692, FLUSSER20001405,10.1007/s10916-017-0867-4, Zhang2015PathologicalBD,e4daac1d08c148ea90ba2590ef20184f}.
ZMs are a family of orthogonal moments due to the property of Zernike polynomial functions \citep{ref:mukundan2001}. Due to the exponential phase term of complex Zernike polynomials, the magnitude of ZMs is rotation invariant. 
In the literature, a comprehensive review of Zernike polynomials and applications were explained \citep{MUKUNDAN19951433, BELKASIM1996577, ZHENJIANG2000169, GU20022905, CHONG2003731, SIM2004331, MITZIAS2004315, PAPAKOSTAS2006960, PAPAKOSTAS20072802, Sadeghi_2021, Niu_2022, Capalbo_2022}.  

Recently, ZMs have been widely used for describing the characteristics of various digital images in different disciplines \citep{GU20022905, CHONG2003731, SIM2004331, MITZIAS2004315, PAPAKOSTAS2006960, PAPAKOSTAS20072802, Sadeghi_2021, Niu_2022, Capalbo_2022}.

The ZMs, as a basis of machine learning, were applied for the identification of solar small-scale brightenings \citep{Yousefzadeh2016, Javaherian2014, Shokri2022, HosseiniRad2021} and small-scale (mini) dimmings \citep{alipour2012,Honarbakhsh2016,Alipour2022}. The ZMs are valuable features for classifying solar flaring and non-flaring active regions \citep{2016cosp...41E1618R,wheatland2017prediction,alipour2019} that developed a tool of solar flare for casting.         

The layout this paper is: Sections \ref{sec:Zp} and \ref{sec:ZM}
describe the Zernike polynomials and Zernike moments, respectively. Section \ref{sec:method} provides an overview of Python code. Section \ref{sec:con} gives the conclusions.

\section{Zernike polynomials}\label{sec:Zp}
The ZPs are a complete set of orthogonal continuous functions in a unit disk. The even ZPs with order $n$ and repetition $m$ in the polar coordinate are given by 

\begin{equation}
    ZP_{pq}(r,\theta) = R_{pq}(r)\,\cos(q\,\theta)
\end{equation}

and the odd ZPs function is defined by
\begin{equation}
    ZP_{p-q}(r,\theta) = R_{pq}(r)\,\sin(q\,\theta)
\end{equation}

where the radial distance in a unit circle is $0\le r \le1$ and the azimuths angle is $0\le \theta \le 2\pi$. 

The radial polynomials for a given set of $p$ and $q$ are defined by

\begin{equation}
     R_{pq}(r) = \sum_{k=0}^{\tfrac{p-q}{2}} \frac{(-1)^k\,(p-k)!}{k!\left (\tfrac{p+q}{2}-k \right )! \left (\tfrac{p-q}{2}-k \right)!} r^{p-2k}
     \label{Vpq}
\end{equation}
 
in which  $p-q=$ even and $|q|\le p$.  The ZPs satisfy the following orthogonality property as, 

 \begin{eqnarray}
 \int _{0} ^{2\pi}{\int _{0} ^{1}  V^{*}_{pq} V_{p'q'} r dr d\theta} = \frac{\pi}{p+1} \delta_{pp'}\delta_{qq'},
  \end{eqnarray}
where $ \delta $ indicates the Kronecker delta function and $V_{pq}$ is the Zernike polynomials.

\begin{figure} 
 \centerline{\includegraphics[width=10cm]{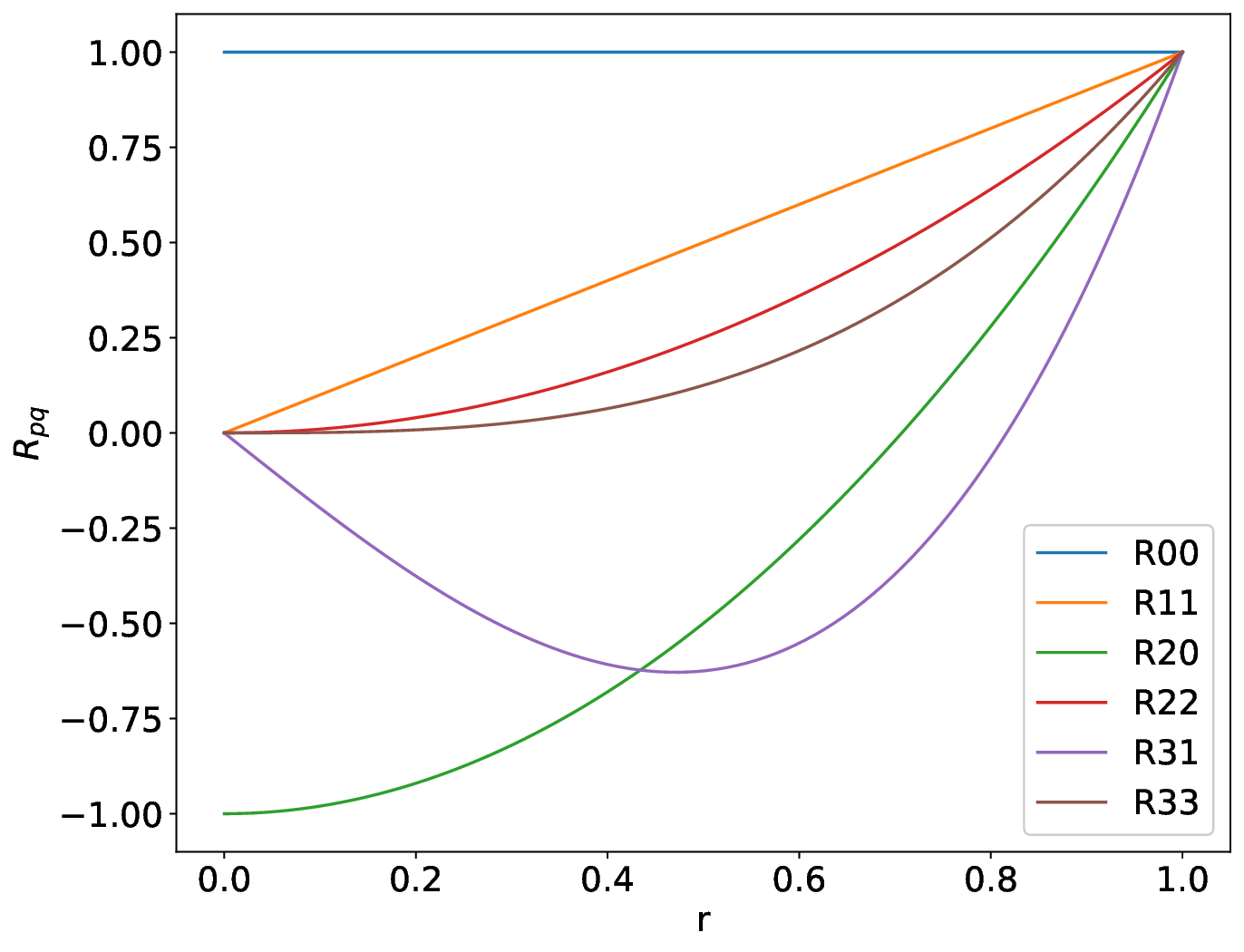}}
 \caption[]{The radial function $R_{pq}$ for a set of $p$ and $q$ in versus radial distance of polar coordinates.}
 \label{radial}
\end{figure}

Figure \ref{radial} represents the radial function $R_{pq}$ for a set of $p$ and $q$ in versus radial distance of polar coordinate. We observe that the radial functions increase oscillations by increasing the order number $p$. This property of ZPs' radial functions is one of the main reasons for applying the ZMs to describe an image in a polar coordinate.  

\begin{figure} 
 \centerline{\includegraphics[width=15cm]{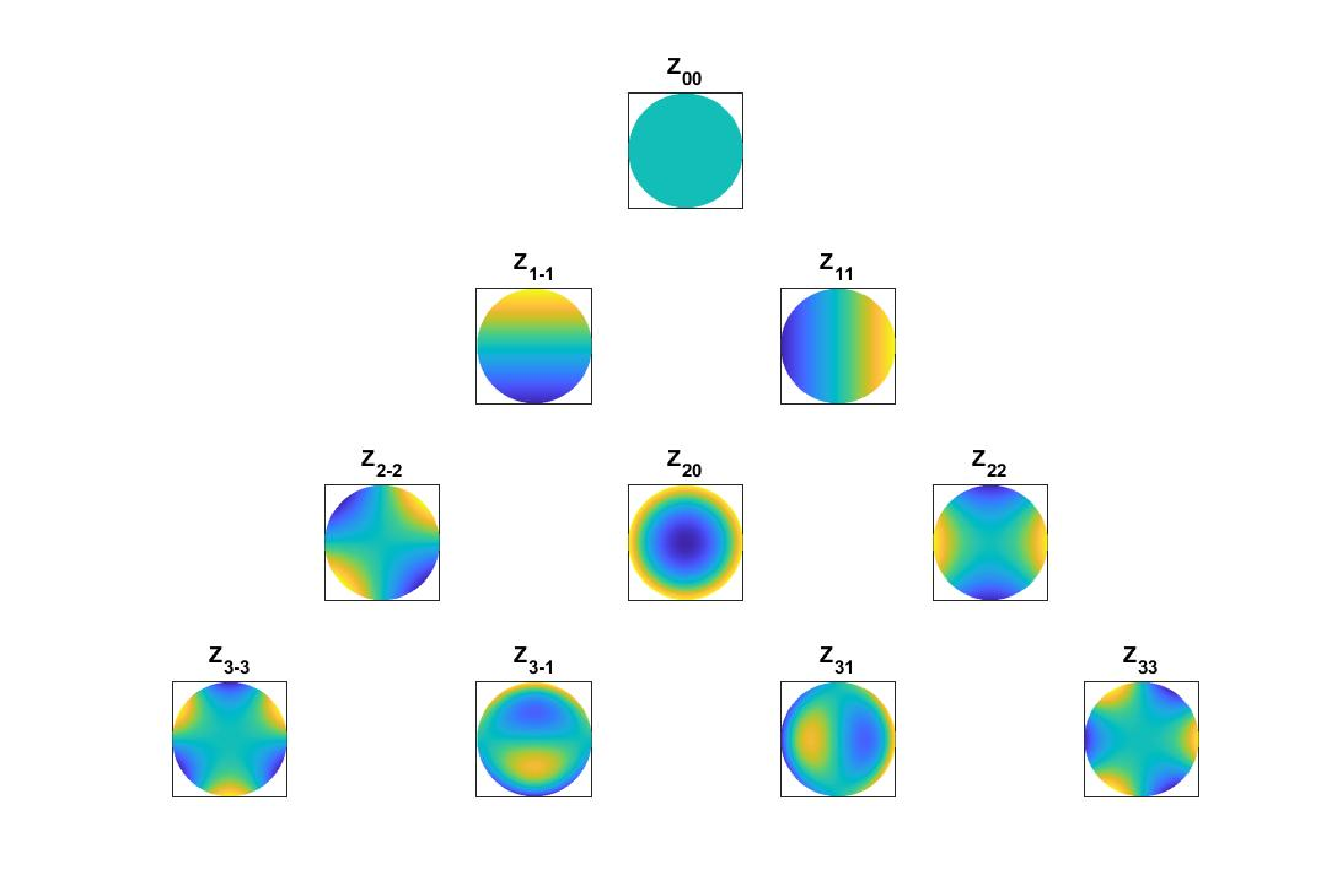}}
 \caption[]{The Zernike polynomial $Z_{pq}$ for a set of order number $p=$ 0 (first row), $p=$1 (second row), $p=$2 (third row), and $p=$3 (fourth row). }
 \label{zp}
\end{figure}

Figure \ref{zp} displays $Z_{pq}$ for a set of order number $p=0,~1,~2,$ and 3 in polar coordinates. The figure shows that each Zernike polynomials have unique radial and azimuthal structures in polar coordinates. This essential characteristic of Zernike polynomials is the main reason for describing an image based on the set of complex Zernike polynomials (combination of even and odd Zernike functions in complex number plane), which maps to a unit circle.

\section{Zernike Moments}\label{sec:ZM}
The reason to describe an image by a set of functions is due to The uniqueness theorem. This theorem explains that the moments are uniquely discriminated for a given image  \citep{ref:hu1962visual}. Contrariwise, we can reconstruct the original image using the set of moments. Moments can specify the properties, such as the centroid of an image, its orientation, and the geometry of the objects. Raw and central moments were defined by  \cite{ref:hu1962visual, ref:grubbstrom2006moments}.

The Zernike moments (ZMs) express an image in a set of complex numbers using the Zernike polynomials $V_{pq}=\rm ZP_{pq}({\rm even})+i\rm ZP_{pq}({\rm odd})$ \citep{ref:mukundan2001}. The image coordinates $(x,y)$ must be transformed into the polar coordinate. The circle's center in polar coordinates is the centroid of an image. For an image function $ I(r,\theta) $, the ZM is given by,

\begin{eqnarray}
 Z_{pq}=\frac{p+1}{\pi}\int _{0} ^{2\pi} {\int _{0} ^{1} {I(r,\theta) V^{*}_{pq} rdrd\theta}}.
 \label{Zpq}
\end{eqnarray}

For a digital image with M $\times$ N pixels, the ZMs are introduced by 
\begin{eqnarray}
 Z_{pq}=\frac{p+1}{\pi}\sum _{i=0} ^{M-1} {\sum _{j=0} ^{N-1} {I(i,j) }R_{pq}(r_{ij}) \exp(-ip \theta_{ij})},
\end{eqnarray}
 where $ r_{ij}=\sqrt{x_{i}^{2}+y_{j}^{2}} $ and $ \theta_{ij}=\arctan(\frac{y_{j}}{x_{i}})  $ are the image cell mapped to a unit disk \citep{ref:wolf2011}. 
 
The Zernike moment array includes elements for a set of order $p$=0 to a maximum order number $P_{\rm max}$. So, the length of Zernike moments 
 $(NZMs)$ is introduced by \citep{alipour2019}
  
\begin{eqnarray}
    NZMs=\sum^{P_{\rm max}}_{p=0} (p+1).
     \label{pmax}
\end{eqnarray}

The reconstructed image ($ I_{R} $) is given by an inverse transformation \cite{ref:khotanzad1990} as follow, 

\begin{eqnarray}
   I_{R}(r,\theta)=\sum _{p=0}^{P_{\rm max}} \sum _{q} {Z_{pq}V_{pq}(r,\theta)}.
\end{eqnarray}
Using the original and reconstructed images, we obtain the reconstruction error as
  
  \begin{eqnarray}
  \rm e^{2}(I,I_{R})=\frac{\sum_{i=0}^{M-1}\sum_{j=0}^{N-1} (I(i,j)-I_{R}(i,j))^{2}}{\sum_{i=0}^{M-1}\sum_{j=0}^{N-1} (I(i,j))^{2}}.
  \label{error}
 \end{eqnarray}

\begin{figure} 
 \centerline{\includegraphics[width=15cm]{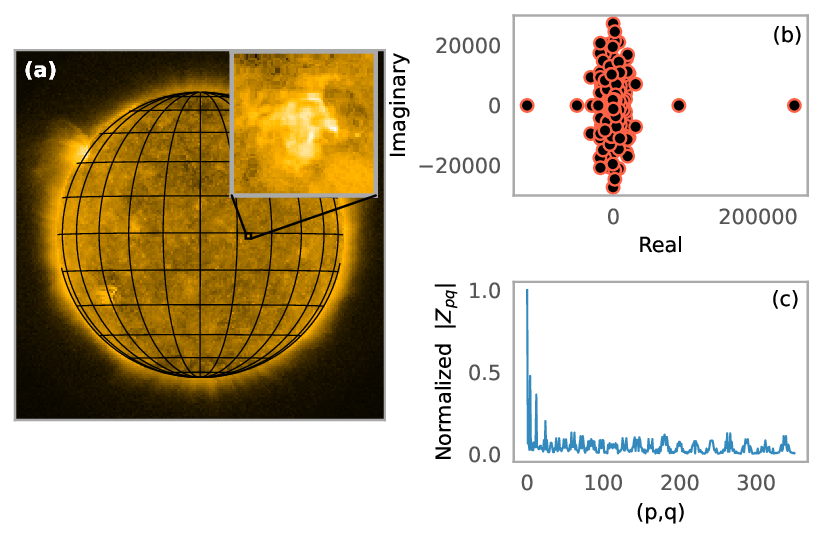}}
 \caption[]{The full disk AIA image at 171 \AA\ that inset a solar coronal bright point (a), the imaginary  and real parts of Zernike moments for a maximum order number of 25 (b), and the absolute normalized Zernike moments versus labels $(p,q)$ (c). }
 \label{AIA}
\end{figure}

Figure \ref{AIA} shows a full disk AIA image at 171 \AA\ inset a solar coronal bright point and the Zernike moments' maximum order 25. The Zernike moments include the imaginary and real parts (panel b). The structure of the moment series is represented by the absolute normalized Zernike moments versus labels. 

\begin{figure} 
 \centerline{\includegraphics[width=14cm]{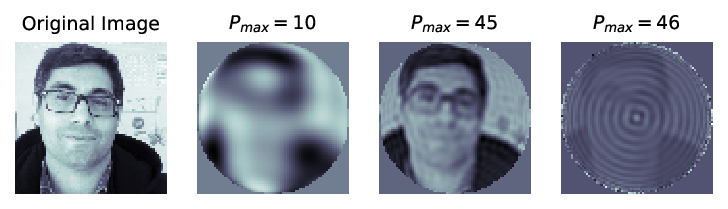}}
 \caption[]{From left to right panels represent the original (face: Hossein Safari) image and reconstructed images with the different maximum order numbers ($P_{\max}$= 10, 45, and 46), respectively.}
 \label{HS}
\end{figure}

Figure \ref{HS} represents an original (face) image and reconstructed images with different maximum order numbers. We observe that the reconstructed image at $P_{\max}$= 10 deviated from the original image, but the reconstructed image at 45 well matched the original image. Also, increasing the maximum order number of the reconstructed image showed noisy image may be due to the discrete behavior of a digital image.

\begin{figure}
\centering
\hspace*{1cm}
\includegraphics[width=\linewidth]{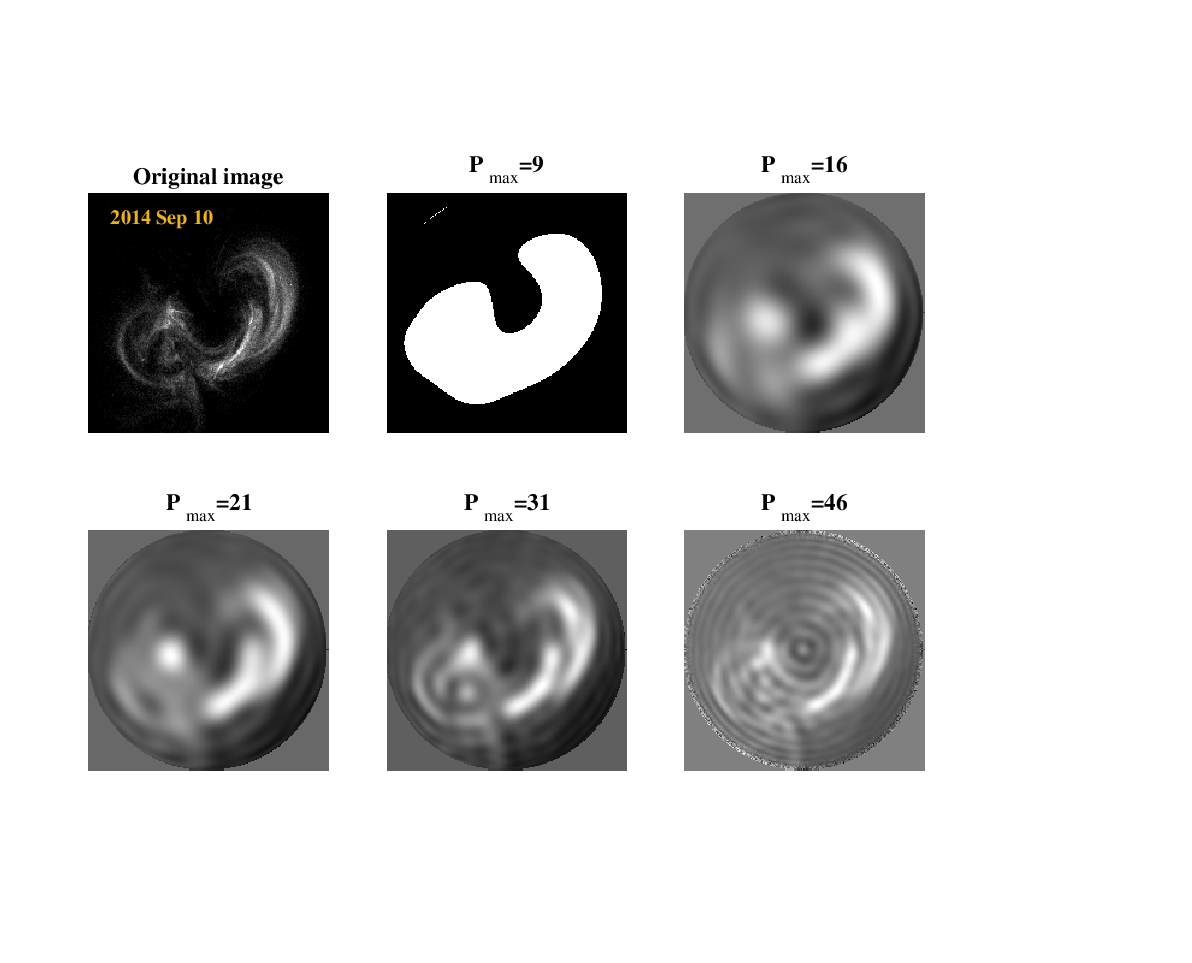}
\vspace*{-2.5cm}
       \caption{ An active region  (sigmiod: left top panel) from SDO/AIA at 94 \AA. The reconstructed images for $P_{\max}$= 9, 16, 21, 31, and 46 \citep{alipour2019}.  }
            \label{fig3}
\end{figure}

Figure \ref{fig3} displays a solar active region (AR) in Solar Dynamics Observatory/Atmospheric Imaging Assembly  (SDO/AIA) at 94 \AA. An sigmiod event and the reconstructed images with various maximum order numbers ($ P_{\rm max}$) are shown. For small $ P_{\rm max} (<10) $, the overall shape of the sigmiod  was reconstructed. We observe that with increasing the $ P_{\rm max}$, the reconstructed image approaches the original image at $ P_{\rm max}~ (=31)$. We also see that the reconstructed image deviates from the original for large $ P_{\rm max}$ (= 46). 


\begin{figure}
     \centering
      \begin{subfigure}
         \centering
         \includegraphics[width=13cm,height=3.8cm]{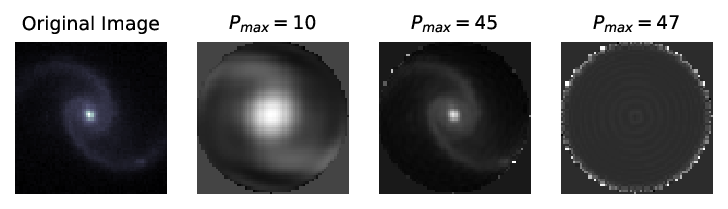}
     \end{subfigure}\\
     \begin{subfigure}
         \centering
         \includegraphics[width=13cm,height=3.8cm]{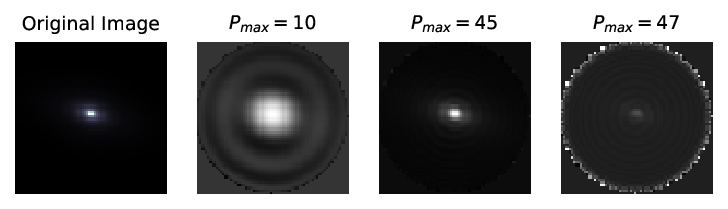}
      \end{subfigure}\\
     \begin{subfigure}
         \centering
         \includegraphics[width=13cm,height=3.8cm]{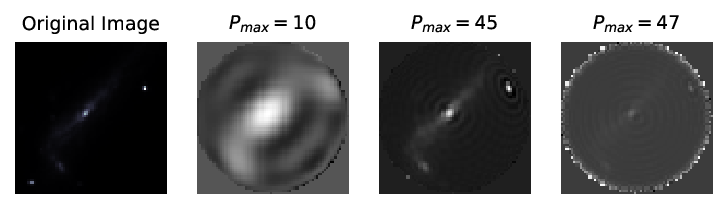}
     \end{subfigure}\\
      \caption{ From left to right, panels represent the original and reconstructed images with the different maximum order numbers ($P_{\max}$= 10, 45, and 47), respectively, for a spiral galaxy (top row), elliptical galaxy (middle row), and irregular galaxy (bottom row). Recorded by SDSS survey.}
\label{GS}  
\end{figure}

Figure \ref{GS} shows the original and reconstructed images with the different maximum order numbers for a spiral galaxy (top row), elliptical galaxy (middle row), and irregular galaxy (bottom row). We find the minimum reconstruction error for $P_{\max}$= 45 for spiral, elliptical, and irregular galaxies. For more or less value than 45, the reconstructed image deviated from the original image. In the case of minimal reconstruction error, we expect to well match objects, shapes, and their orientations or morphologies in reconstructed images and original images.  

\begin{figure}
\centering
\includegraphics[width=\linewidth]{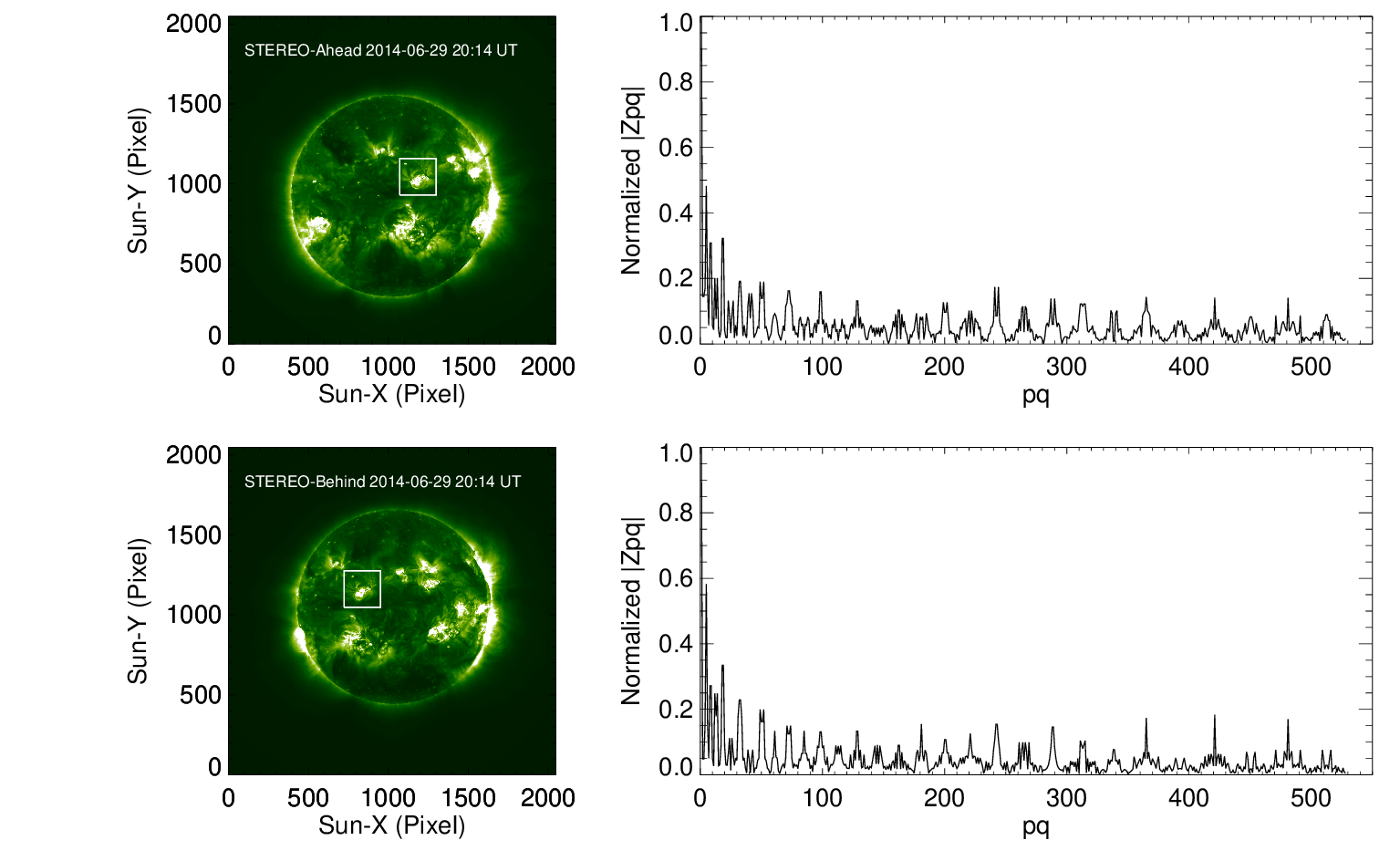}
       \caption{ A solar active region indicated by a white box of the EUVI  images at 195 \AA~ recorded by STEREO-A (Left top) and STEREO-B (Left bottom). The normalized absolute values of the Zernike moments for $P_{\max}$=31 \citep{alipour2019}. }
       \label{fig4}
\end{figure}

ZPs include orthogonal functions; hence moments give the properties of an image. Due to the Fourier term in the azimuthal angle function, the absolute value of moments is independent of the objects' rotation angle in the image.
Space missions and ground base instruments observe solar features from various perspectives and scales. The Soho was in the first Lagrangian point of the Sun-Earth. STEREO A and B are in Earth's orbit. Figure \ref{fig4} presents the ZMs of an active region observed by two STEREO A and B. The ZMs are similar from two different viewpoints. The block structures of the ZMs series are identical, with slight differences. These trivial differences may be due to the digital rather than the continuous image. Applying a transformation (to the image centre of brightness) and image normalization, ZMs will be translation and scaling invariances, respectively \citep[see, e.g.,][]{ref:khotanzad1990}

\begin{figure}
\centering
\includegraphics[width=14cm]{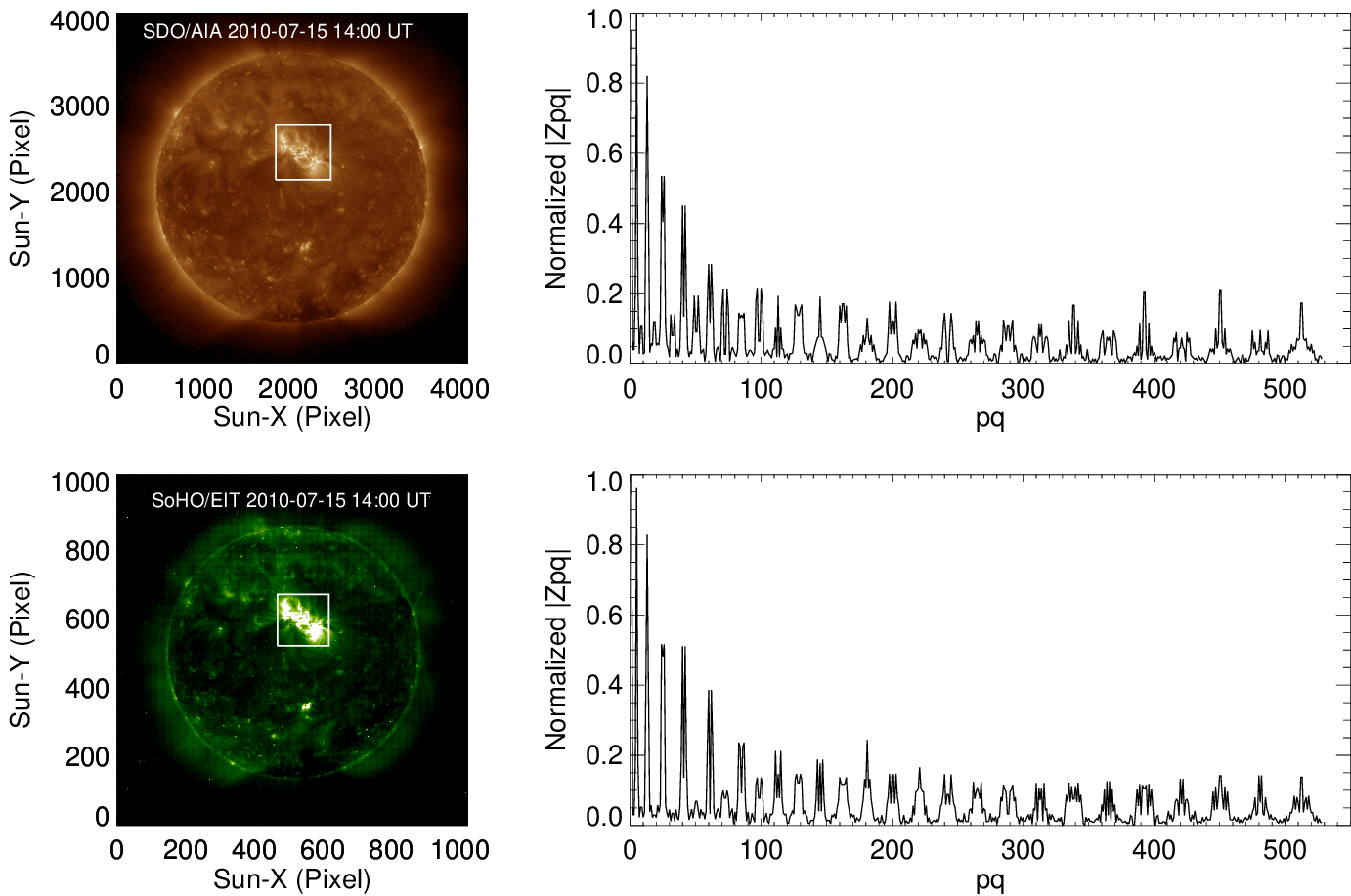}
       \caption{ The white boxes represent of an active region  observed by SDO/AIA image at 193 \AA\ (Left top panel) and SoHO/EIT (Left bottom panel).(Right) The normalized absolute value of Zernike moments for two SDO and SoHo views \citep{alipour2019}.}
       \label{fig5}
\end{figure}

The  SoHO/EIT and SDO/AIA resolutions are 0.6 and 2.4, respectively. The ZMs for the active region  (Figure \ref{fig5}) with various resolutions are slightly similar. It seems the ZMs are functions of the morphology and geometry of the objects and depend less on the object's size.

------------------------------------

\section{Python code for ZMs}\label{sec:method}
The Python code is available at Github (\url{https://github.com/hmddev1/ZEMO}) and PyPI (\url{https://pypi.org/project/ZEMO/1.0.0/}). The Python code calculates ZMs for a given image. \cite{alipour2015} and \cite{alipour2019} used the primitive code for calculating ZMs of solar features. 
The code includes the following functions:

\begin{itemize}

\item The zernike\_order\_list function calculates factorials, $p$ (order numbers)-indices, and $q$ (repetition numbers)-indices for a given maximum order number of Zernike polynomials.

\item The robust\_fact\_quot function removes common elements from lists and calculates product quotients.

\item The zernike\_bf function generates Zernike basis functions stored in a complex-valued grid.

\item The zernike\_mom function calculates Zernike moments by summing the product of the image and basis functions.

\item The zernike\_rec function reconstructs an image by summing the weighted Zernike basis functions via ZMs.
\end{itemize}

The code includes checks for data validity, such as square image size matching, and prints informative error messages.

\section{Conclusion}\label{sec:con}
The Zernike polynomials indicate the distance along the radius and azimuthal angle. Equation (\ref{Zpq}) shows the image function weighted by the radial part $r  R_{pq}(r)$. We note that $|R_{pq}(r)|<1$ and $|r R_{pq}(r)|<r$ within a unit circle that shows that the edge's pixels have more extensive weights than the center pixels. The higher-order Zernike polynomials will show more oscillations to extract information on the image details along the radius from the origin to the perimeter \citep{ref:shutler2001}.

Why are ZMs helpful in expressing an image? 

\begin{itemize}
\item The Zernike basis is orthogonal and complete set functions, so ZMs are unique quantities features. 
\item We may reconstruct the original image by a finite number of moments.  
\item ZMs are slightly sensitive to noises. 
\item The magnitude of ZMs is rotation invariant. Image normalization makes translation and scale invariants for ZMs.
\end{itemize}These reasons showed the capability of ZMs to describe an image to apply machine learning to identify and track the features in several disciplines. We published the Python code via Github and PyPI.

\clearpage
\bibliographystyle{apalike}
\bibliography{MS.bib}

\begin{thebibliography}{}

\bibitem[{Alipour} et~al., 2019]{alipour2019}
{Alipour}, N., {Mohammadi}, F., and {Safari}, H. (2019).
\newblock {Prediction of Flares within 10 Days before They Occur on the Sun}.
\newblock {\em \apjs}, 243(2):20.

\bibitem[Alipour and Safari, 2015]{alipour2015}
Alipour, N. and Safari, H. (2015).
\newblock Statistical properties of solar coronal bright points.
\newblock {\em The Astrophysical Journal}, 807(2):175.

\bibitem[{Alipour} et~al., 2012]{alipour2012}
{Alipour}, N., {Safari}, H., and {Innes}, D.~E. (2012).
\newblock {An Automatic Detection Method for Extreme-ultraviolet Dimmings
  Associated with Small-scale Eruption}.
\newblock {\em \apj}, 746:12.

\bibitem[{Alipour} et~al., 2022]{Alipour2022}
{Alipour}, N., {Safari}, H., {Verbeeck}, C., {Berghmans}, D., {Auch{\`e}re},
  F., {Chitta}, L.~P., {Antolin}, P., {Barczynski}, K., {Buchlin}, {\'E}.,
  {Aznar Cuadrado}, R., {Dolla}, L., {Georgoulis}, M.~K., {Gissot}, S.,
  {Harra}, L., {Katsiyannis}, A.~C., {Long}, D.~M., {Mandal}, S., {Parenti},
  S., {Podladchikova}, O., {Petrova}, E., {Soubri{\'e}}, {\'E}., {Sch{\"u}hle},
  U., {Schwanitz}, C., {Teriaca}, L., {West}, M.~J., and {Zhukov}, A.~N.
  (2022).
\newblock {Automatic detection of small-scale EUV brightenings observed by the
  Solar Orbiter/EUI}.
\newblock {\em \aap}, 663:A128.

\bibitem[Aschwanden, 2010]{ref:aschwanden2010image}
Aschwanden, M.~J. (2010).
\newblock Image processing techniques and feature recognition in solar physics.
\newblock {\em Solar Physics}, 262(2):235--275.

\bibitem[Belkasim et~al., 1996]{BELKASIM1996577}
Belkasim, S., Ahmadi, M., and Shridhar, M. (1996).
\newblock Efficient algorithm for fast computation of zernike moments.
\newblock {\em Journal of the Franklin Institute}, 333(4):577--581.

\bibitem[Capalbo et~al., 2022]{Capalbo_2022}
Capalbo, V., Petris, M.~D., Luca, F.~D., Cui, W., Yepes, G., Knebe, A., Rasia,
  E., Ruppin, F., and Ferragamo, A. (2022).
\newblock Morphological analysis of {SZ} and x-ray maps of galaxy clusters with
  zernike polynomials.
\newblock {\em {EPJ} Web of Conferences}, 257:00008.

\bibitem[Chong et~al., 2003]{CHONG2003731}
Chong, C.-W., Raveendran, P., and Mukundan, R. (2003).
\newblock A comparative analysis of algorithms for fast computation of zernike
  moments.
\newblock {\em Pattern Recognition}, 36(3):731--742.

\bibitem[Doerr and Florence, 2020]{e4daac1d08c148ea90ba2590ef20184f}
Doerr, F. and Florence, A. (2020).
\newblock A micro-xrt image analysis and machine learning methodology for the
  characterisation of multi- particulate capsule formulations.
\newblock {\em International Journal of Pharmaceutics: X}, 2.

\bibitem[Flusser, 2000]{FLUSSER20001405}
Flusser, J. (2000).
\newblock On the independence of rotation moment invariants.
\newblock {\em Pattern Recognition}, 33(9):1405--1410.

\bibitem[{Gonzaga} and {Ferreira Costa}, 1996]{ref:hu1962visual}
{Gonzaga}, A. and {Ferreira Costa}, J.~A. (1996).
\newblock {Moment invariants applied to the recognition of objects using neural
  networks}.
\newblock In {Tescher}, A.~G., editor, {\em Applications of Digital Image
  Processing XIX}, volume 2847 of {\em Society of Photo-Optical Instrumentation
  Engineers (SPIE) Conference Series}, pages 223--233.

\bibitem[Gonzalez and Faisal, 2019]{book}
Gonzalez, R. and Faisal, Z. (2019).
\newblock {\em Digital Image Processing Second Edition}.

\bibitem[Goshtasby, 1985]{ref:goshtasby1985}
Goshtasby, A. (1985).
\newblock Template matching in rotated images.
\newblock {\em IEEE Transactions on Pattern Analysis and Machine Intelligence},
  (3):338--344.

\bibitem[Grubbstr{\"o}m and Tang, 2006]{ref:grubbstrom2006moments}
Grubbstr{\"o}m, R.~W. and Tang, O. (2006).
\newblock The moments and central moments of a compound distribution.
\newblock {\em European Journal of Operational Research}, 170(1):106--119.

\bibitem[Gu et~al., 2002]{GU20022905}
Gu, J., Shu, H., Toumoulin, C., and Luo, L. (2002).
\newblock A novel algorithm for fast computation of zernike moments.
\newblock {\em Pattern Recognition}, 35(12):2905--2911.
\newblock Pattern Recognition in Information Systems.

\bibitem[Heywood and Noakes, 1995]{ref:heywood1995}
Heywood, M. and Noakes, P. (1995).
\newblock Fractional central moment method for movement-invariant object
  classification.
\newblock {\em IEE Proceedings-Vision, Image and Signal Processing},
  142(4):213--219.

\bibitem[{Honarbakhsh} et~al., 2016]{Honarbakhsh2016}
{Honarbakhsh}, L., {Alipour}, N., and {Safari}, H. (2016).
\newblock {Magnetic Evolution of Mini-Coronal Mass Ejections}.
\newblock {\em Solar Physics}, 291(3):941--952.

\bibitem[{Honarbakhsh} and {Morra}, 2023]{honarbakhsh}
{Honarbakhsh}, L. and {Morra}, G. (2023).
\newblock {Classification of IR Images of Small Eruptions at the Erebus
  Volcano, Antarctica, With Zernike Moments and Support Vector Machine}.
\newblock {\em Journal of Geophysical Research (Solid Earth)},
  128(6):e2022JB025728.

\bibitem[{Hosseini Rad} et~al., 2021]{HosseiniRad2021}
{Hosseini Rad}, S., {Alipour}, N., and {Safari}, H. (2021).
\newblock {Energetics of Solar Coronal Bright Points}.
\newblock {\em \apj}, 906(1):59.

\bibitem[Hu, 1962]{1057692}
Hu, M.-K. (1962).
\newblock Visual pattern recognition by moment invariants.
\newblock {\em IRE Transactions on Information Theory}, 8(2):179--187.

\bibitem[Javaherian et~al., 2014]{Javaherian2014}
Javaherian, M., Safari, H., Amiri, A., and Ziaei, S. (2014).
\newblock Automatic method for identifying photospheric bright points and
  granules observed by sunrise.
\newblock {\em Solar Physics}, 289.

\bibitem[Khotanzad and Hong, 1990]{ref:khotanzad1990}
Khotanzad, A. and Hong, Y.~H. (1990).
\newblock Invariant image recognition by zernike moments.
\newblock {\em IEEE Transactions on pattern analysis and machine intelligence},
  12(5):489--497.

\bibitem[Mitzias and Mertzios, 2004]{MITZIAS2004315}
Mitzias, D.~A. and Mertzios, B.~G. (2004).
\newblock A neural multiclassifier system for object recognition in robotic
  vision applications.
\newblock {\em Measurement}, 36(3-4):315--330.

\bibitem[Moradkhani et~al., 2015]{MORADKHANI2015123}
Moradkhani, M., Amiri, A., Javaherian, M., and Safari, H. (2015).
\newblock A hybrid algorithm for feature subset selection in high-dimensional
  datasets using fica and iwssr algorithm.
\newblock {\em Applied Soft Computing}, 35:123--135.

\bibitem[Mukundan et~al., 2001]{ref:mukundan2001}
Mukundan, R., Ong, S., and Lee, P.~A. (2001).
\newblock Image analysis by tchebichef moments.
\newblock {\em IEEE Transactions on image Processing}, 10(9):1357--1364.

\bibitem[Mukundan and Ramakrishnan, 1995]{MUKUNDAN19951433}
Mukundan, R. and Ramakrishnan, K.~R. (1995).
\newblock Fast computation of legendre and zernike moments.
\newblock {\em Pattern Recognit.}, 28:1433--1442.

\bibitem[Nayak et~al., 2018]{10.1007/s10916-017-0867-4}
Nayak, D.~R., Dash, R., and Majhi, B. (2018).
\newblock An improved pathological brain detection system based on
  two-dimensional pca and evolutionary extreme learning machine.
\newblock {\em J. Med. Syst.}, 42(1):1–15.

\bibitem[{Niu} and {Tian}, 2022]{Niu_2022}
{Niu}, K. and {Tian}, C. (2022).
\newblock {Zernike polynomials and their applications}.
\newblock {\em Journal of Optics}, 24(12):123001.

\bibitem[{Noori} et~al., 2019]{Noori2019}
{Noori}, M., {Javaherian}, M., {Safari}, H., and {Nadjari}, H. (2019).
\newblock {Statistics of photospheric supergranular cells observed by SDO/HMI}.
\newblock {\em Advances in Space Research}, 64(2):504--513.

\bibitem[Papakostas et~al., 2007]{PAPAKOSTAS20072802}
Papakostas, G.~A., Boutalis, Y.~S., Karras, D.~A., and Mertzios, B.~G. (2007).
\newblock A new class of zernike moments for computer vision applications.
\newblock {\em Information Sciences}, 177(13):2802--2819.

\bibitem[Papakostas et~al., 2006]{PAPAKOSTAS2006960}
Papakostas, G.~A., Boutalis, Y.~S., Papaodysseus, C., and Fragoulis, D.~K.
  (2006).
\newblock Numerical error analysis in zernike moments computation.
\newblock {\em Image and Vision Computing}, 24(9):960--969.

\bibitem[{Raboonik} et~al., 2016]{2016cosp...41E1618R}
{Raboonik}, A., {Safari}, H., {Dadashi}, N., and {Alipour}, N. (2016).
\newblock {Prediction of M and X Solar flares by Using Machine Learning
  Algorithm}.
\newblock In {\em 41st COSPAR Scientific Assembly}, volume~41, pages
  D2.5--21--16.

\bibitem[Sadeghi et~al., 2021]{Sadeghi_2021}
Sadeghi, M., Javaherian, M., and Miraghaei, H. (2021).
\newblock Morphological-based classifications of radio galaxies using
  supervised machine-learning methods associated with image moments.
\newblock {\em The Astronomical Journal}, 161(2):94.

\bibitem[Shokri et~al., 2022]{Shokri2022}
Shokri, Z., Alipour, N., Safari, H., Kayshap, P., Podladchikova, O., Nigro, G.,
  and Tripathi, D. (2022).
\newblock Synchronization of small-scale magnetic features, blinkers, and
  coronal bright points.
\newblock {\em The Astrophysical Journal}, 926:42.

\bibitem[Shutler and Nixon, 2001]{ref:shutler2001}
Shutler, J.~D. and Nixon, M.~S. (2001).
\newblock Zernike velocity moments for description and recognition of moving
  shapes.
\newblock In {\em Proceedings of the British Machine Vision Conference}, pages
  72.1--72.10. BMVA Press.
\newblock doi:10.5244/C.15.72.

\bibitem[Sim et~al., 2004]{SIM2004331}
Sim, D.-G., Kim, H.-K., and Park, R.-H. (2004).
\newblock Invariant texture retrieval using modified zernike moments.
\newblock {\em Image and Vision Computing}, 22(4):331--342.

\bibitem[Teh and Chin, 1988]{ref:teh1988image}
Teh, C.-H. and Chin, R.~T. (1988).
\newblock On image analysis by the methods of moments.
\newblock In {\em Computer Vision and Pattern Recognition, 1988. Proceedings
  CVPR'88., Computer Society Conference on}, pages 556--561. IEEE.

\bibitem[Wheatland et~al., 2017]{wheatland2017prediction}
Wheatland, M., Alipour, N., Raboonik, A., and Safari, H. (2017).
\newblock Prediction of solar flares using unique signatures of magnetic field
  images.

\bibitem[Wolf et~al., 2011]{ref:wolf2011}
Wolf, C., Taylor, G., Jolion, J.-M., Verne, B.~J., and Einstein, A.~A. (2011).
\newblock Learning individual human activities from short binary shape
  sequences.
\newblock Technical report, Technical Report LIRIS. Available at http://liris.
  cnrs. fr/Documents/Liris-5294. pdf.

\bibitem[Yousefzadeh et~al., 2016]{Yousefzadeh2016}
Yousefzadeh, M., Safari, H., Attie, R., and Alipour, N. (2016).
\newblock Motion and magnetic flux changes of coronal bright points relative to
  supergranular cell boundaries.
\newblock {\em Solar Physics}, 291.

\bibitem[Zhang et~al., 2015]{Zhang2015PathologicalBD}
Zhang, Y., Wang, S., Sun, P., and Phillips, P. (2015).
\newblock Pathological brain detection based on wavelet entropy and hu moment
  invariants.
\newblock {\em Bio-medical materials and engineering}, 26 Suppl 1:S1283--90.

\bibitem[Zheng et~al., 2015]{ref:zheng2015improved}
Zheng, C., Pulido, J., Thorman, P., and Hamann, B. (2015).
\newblock An improved method for object detection in astronomical images.
\newblock {\em Monthly Notices of the Royal Astronomical Society},
  451(4):4445--4459.

\bibitem[{Zhenjiang}, 2000]{ZHENJIANG2000169}
{Zhenjiang}, M. (2000).
\newblock {Zernike moment-based image shape analysis and its application}.
\newblock {\em Pattern Recognition Letters}, 21(2):169--177.

\end{thebibliography}


\end{document}